# Quantum Hall States of High Symmetry

CHETAN NAYAK[*]

*Department of Physics*

*Joseph Henry Laboratories*

*Princeton University*

*Princeton, N.J. 08544*

FRANK WILCZEK[†]

*School of Natural Sciences*

*Institute for Advanced Study*

*Olden Lane*

*Princeton, N.J. 08540*

---

[*] Research supported in part by a Fannie and John Hertz Foundation fellowship. nayak@puhep1.princeton.edu

[†] Research supported in part by DOE grant DE-FG02-90ER40542. WILCZEK@IASSNS.BITNET

# ABSTRACT

We identify some hidden symmetries of Chern-Simons theories, such as appear in the effective theory for quantized Hall states. This allows us to determine which filling fractions admit spin-singlet quantum Hall states. Our results shed some light on states already observed at $\nu = 2/3$, and transitions between them. We identify $SU(2)$, or higher, symmetries of many additional states – including spin-polarized states. Our symmetries classify low-lying excited states and may be of use in the construction of trial wavefunctions, but are typically not present in the edge theory, where they are lifted by non-universal couplings.

# 1. Introduction

In this paper we shall discuss some hidden symmetries of certain effective Chern-Simons gauge theories that govern incompressible quantized Hall states. These symmetries have immediate implications for experimental and numerical studies of the states which exhibit them. We shall briefly conjecture on some wider implications at the end.

Several years ago Halperin [1] pointed out that band mass and $g$ factor corrections make the ratio of Zeeman to cyclotron energies $\sim 7/400$, so that it may be necessary to include both spins when describing electrons in moderately strong magnetic fields, even when the filling fraction is less than unity. As a specific example, he introduced the $(m, m, n)$ wavefunctions:

$$\Psi_{(m,m,n)}(w_i, z_j) = \prod_{i<j}(w_i - w_j)^m \prod_{i<j}(z_i - z_j)^m \prod_{i,j}(w_i - z_j)^n . \qquad (1.1)$$

where the $w_i$'s and $z_j$'s are, respectively, the coordinates of the up and down spin electrons, and the filling fraction is $\nu = \frac{2}{m+n}$. It appears that these wavefunctions single out a preferred spin quantization axis, and in general they do, but for $n = m - 1$ they describe spin singlets. Thus it is natural to expect spin singlet states at filling fractions $\nu = \frac{2}{2m-1}$, with $m$ odd required by Fermi statistics. And in fact there is good evidence, both experimental and numerical, that such states occur. But without a deeper understanding of the origin of the spin symmetry – in particular, a connection to the universal properties of the correlations (as opposed to a specific trial wave function) – several obvious questions are hard to address, notably: are these the only spin-singlet states? This problem is highlighted by the fact that prime experimental candidates for spin-singlet quantum Hall states are found at $\nu = 2/3$ and $\nu = 4/3$ [2] (see also the numerical results [3,4]) which are not among the $(m, m, m-1)$ states or the hierarchy built on them.

To address it, we shall exploit the formalism proposed by Wen and Zee [5], in which a matrix $K$ determines the universal long-range correlations of a Hall fluid, independent of any particular construction of trial wavefunctions. We shall explicitly construct candidate symmetry generators – related to the holonomies of gauge fields – directly within the effective field theory, and thus identify the conditions that a $K$-matrix must satisfy in order to describe a spin-singlet state. Our criterion naturally leads us, among other things, to $K$-matrices for the $\nu = 2/3, 4/3$ states. The trial wavefunction of Wu, Dev, and Jain [7] for the $\nu = 2/3$ state plausibly falls in the universality class described by the proposed $K$-matrix. This $K$-matrix is equivalent, by a change of basis, to that of the spin-polarized $\nu = 2/3$ state, a fact that has interesting implications for the transition between spin-singlet and polarized states in tilted-field experiments.

Multi-component quantum Hall fluids also arise in double-layer and single wide layer systems and, of course, in the original hierarchy of polarized, single-layer Hall systems. Our results have implications for these systems, as well. Certain hierarchical states, in particular, are distinguished by large symmetry groups that may be constructed in the effective field theory. While these symmetries are generically broken by non-universal edge effects, they provide selection rules for processes in the bulk, and may also give a useful means of constructing trial wavefunctions.

## 2. Symmetries of the Effective Field Theory

The long-wavelength Chern-Simons effective theory of an abelian quantum Hall state can be written in the form [5]:

$$L = \frac{1}{4\pi} K_{ij} \epsilon_{\mu\nu\lambda} a^i_\mu \partial_\nu a^j_\lambda + \frac{1}{2\pi} \epsilon_{\mu\nu\lambda} A_\mu t^i \partial_\nu a^i_\lambda + j^i_\mu a^i_\mu + \ldots \qquad (2.1)$$

$j^i_\mu$ is the current of vortices of the $i^{th}$ type; the ellipsis refers to interactions of higher order in gradients, that are irrelevant at long wavelength. This effective field theory is equivalent to the Landau-Ginzburg theory

$$L = (\partial_\mu \phi^i + \alpha^i_\mu + t^i A_\mu)^2 + \frac{1}{2\pi} \alpha^i_\mu \epsilon_{\mu\nu\lambda} \partial_\nu \beta^i_\lambda + \frac{1}{4\pi} K_{ij} \epsilon_{\mu\nu\lambda} \beta^i_\mu \partial_\nu \beta^j_\lambda \, , \qquad (2.2)$$

as may be seen by making the change of variables $\partial_\mu \phi^i = \frac{1}{4\pi} \epsilon_{\mu\nu\lambda} \partial_\nu a^i_\lambda$ and inte-grating out $\alpha^i$. One might also consider a dual theory, by integrating out $\beta^i$. The Hall conductance is given by

$$\sigma_H = \sum_{i,j} t^i t^j (K^{-1})_{ij} \qquad (2.3)$$

while the charge of a vortex (*i.e.* a physical quasiparticle) of type $i$ is:

$$q_i = \sum_j t^j (K^{-1})_{ij} \qquad (2.4)$$

and the braiding statistics between vortices of types $i$ and $j$ is:

$$\theta_{ij} = (K^{-1})_{ij} \ . \qquad (2.5)$$

Implicit in the normalizations (and very important for our purposes) is the assumption that the charges associated with the $j^i_\mu$ are quantized in integers. Distinct quantum Hall states are therefore represented by equivalence classes of $(K,t)$ pairs under $SL(\kappa, Z)$ basis changes, where $\kappa$ is the rank of the $K$-matrix.

Consider first the case of a single Chern-Simons field on a bounded, simply-connected region:

$$L = \frac{k}{4\pi} \epsilon_{\mu\nu\lambda} a_\mu \partial_\nu a_\lambda + \frac{1}{\sqrt{2}} j_\mu a_\mu \ . \qquad (2.6)$$

The $\sqrt{2}$ occurs in the coupling to vortices because the properly normalized $a$ with kinetic term (2.6) will, in the cases of interest, be of the form $a = \frac{1}{\sqrt{2}}(a_1 - a_2)$ with coupling to sources of the form $\frac{1}{\sqrt{2}}(j_1 - j_2)a$ or simply $\frac{1}{\sqrt{2}} ja$ if $j = j_1 - j_2$ ($j$ is still quantized in integers). Let us choose the gauge $a_0 = 0$. Then the Lagrangian is:

$$L = \frac{k}{4\pi}\left(a_2 \frac{\partial}{\partial t} a_1 - a_1 \frac{\partial}{\partial t} a_2\right) + \frac{1}{\sqrt{2}} j_i a_i \qquad (2.7)$$

and the equal-time commutation relations which follow from it are:

$$[a_2(x,t), a_1(x',t)] = \frac{2\pi}{k} \delta(x - x') \qquad (2.8)$$

We must also impose the constraint (the $a_0$ equation of motion):

$$f_{12} = \frac{1}{k\sqrt{2}} 2\pi j_0 . \tag{2.9}$$

In view of (2.8), the constraint must be imposed as a condition on physical states rather than as an operator identity. The appropriate condition reflecting vortex number quantization is then

$$e^{ik\sqrt{2} \int f_{12}} = 1 . \tag{2.10}$$

The integral is over a two-dimensional region; the identity holds for all bounded regions. This identity is consistent with (2.8) for the gauge-invariant commutators of field strengths with closed Wilson lines.

Consider, now, for $k = 1$ the operators

$$O^3 = \frac{1}{2\sqrt{2\pi}} \oint a \tag{2.11}$$

$$O^+ = \frac{1}{2\sqrt{2\pi}} \oint U(x,0) \tag{2.12}$$

$$O^- = \frac{1}{2\sqrt{2\pi}} \oint U(0,x) \tag{2.13}$$

The line integrals are about the boundary of the region, and the holonomy operator, $U(x,0)$ is defined by:

$$U(x,0) = e^{i\sqrt{2}\int_0^x a} \tag{2.14}$$

where 0 is some point in the interior and the path is arbitrary. The constraint (2.10) insures that the integral is path independent. These operators have the commutation relations:

$$[O^3, O^\pm] = \pm O^\pm \tag{2.15}$$

$$[O^+, O^-] = O^3 \tag{2.16}$$

This is an $SU(2)$ algebra.

The first commutation relation follows immediately from (2.8). The second may be most easily obtained by appealing to a mathematical construction (the Frenkel-Kac construction) in a form familiar to physicists from conformal field theory [6]. Since (2.8) is equivalent to the commutation relations for a free scalar field, $a_i = \frac{1}{\sqrt{2}}\partial_i\phi$, we can derive (2.16) by following the steps of the conformal field theory derivation. The appropriate periodicity of the scalar field is enforced by (2.10). Had we attempted a similar construction for $k > 1$, we would have run into the difficulty that the coefficient of the exponential in $O^{\pm}$ required to give a simple commutator would not yield a path-independent holonomy.

Since the gauge field Hamiltonian vanishes, that part of the theory is trivially invariant under the algebra. There are two terms in (2.1) which give a non-vanishing Hamiltonian, namely the couplings to the electromagnetic field and to vortices. This first term vanishes if the gauge field in question is electrically neutral. Thus if $a_\mu^i$ is the gauge field with unit coupling constant, then $t^i$ must vanish for the theory to be $SU(2)$ symmetric. By requiring that the second term be invariant, we determine the appropriate transformation property of the vortex current $j_\mu^i$ (its transformation law can be determined from the Chern-Simons constraint equation locking the density to the flux). The unit vortex and anti-vortex form a spinor.

Let us return to the case of several Chern-Simons fields, (2.1). Suppose we can make an $SL(\kappa, Z)$ change of basis so that $K$ decomposes into a $(\kappa - 1) \times (\kappa - 1)$ matrix and a $1 \times 1$ matrix which is equal to 1 and that this gauge field has the required coupling to quantized vortices. Then, by the above arguments, the theory has an $SU(2)$ symmetry. Since these effective field theories are $SU(2)$ symmetric and have non-degenerate ground states (on the plane or sphere; the degeneracies on higher genus surfaces are not $SU(2)$ degeneracies), these ground states must be $SU(2)$ singlets.

In general one may have several different $SU(2)$ algebras arising in this way, and the question arises whether they may be composed into larger symmetries. A method for doing this follows readily along the lines already drawn. A general notation gets awkward, so we shall content ourselves here with a few examples. A $\kappa \times \kappa$ $K$-matrix with $m \pm 1$s along the diagonal and $m$s elsewhere, $m$ even, will support $SU(\kappa)$ symmetry. The unit $K$ matrix, corresponding to $\kappa$ filled Landau levels, supports $SO(2\kappa)$ symmetry which is broken down to $SU(\kappa)$ by the coupling to the electromagnetic field.

## 3. Examples of $SU(2)$ Symmetric States

To describe spin-singlet states of electrons, we must make sure that our $SU(2)$ symmetry corresponds to spin rotations. The simplest way to identify the spin operator is to couple in a Zeeman term to the Landau-Ginzburg theory, and find out what this term transforms into in the dual theory. In a basis adapted so that the odd-numbered columns in a $\kappa \times \kappa$ $K$ matrix correspond to spin up and the even to spin down, we find that the Zeeman coupling in the symmetric basis (where all the $t^i = 1$) is of the form:

$$L_{\text{Zeeman}} \propto \epsilon_{ij} \partial_i \Big( \sum_{n=1}^{\kappa/2} a_j^{(2n-1)} - a_j^{(2n)} \Big) \tag{3.1}$$

This is of course analogous to the charge coupling term, which has $t^i = 1$ for all $i$; the only difference is that up and down spins enter with opposite signs. The requirement for $SU(2)$ spin symmetry is that the linear combination of gauge fields on the right-hand-side of (3.1) should be an eigenvector of $K$ with unit eigenvalue.

We can immediately see, at the effective field theory level, that Halperin's $(m, m, m-1)$ states are spin singlets. They have $2 \times 2$ $K$-matrices,

$$K = \begin{pmatrix} m & m-1 \\ m-1 & m \end{pmatrix} \tag{3.2}$$

which may be brought to the diagonal form $K = \text{diag}(2m-1, 1)$. How do we know that this is the $K$-matrix corresponding to the $(m, m, m-1)$ state? The simplest way is to observe that the filling factor and quasiparticle charge and statistics is the same. Another way, which is suggested by the form of the wavefunction,

$$\Psi_{(m,m,m-1)} = \Psi_{(1,1,0)} \times \Psi_{(m-1,m-1,m-1)}, \tag{3.3}$$

is to consider the Landau-Ginzburg theory of the $\nu = 2$ spin-singlet state. If one adds another Chern-Simons field which adds $m-1$ flux tubes to each electron then the mean-field condensed ground state of this theory occurs at $\nu = \frac{2}{2m-1}$. The dual theory to this Landau-Ginzburg theory has the $K$-matrix above.

Other $2 \times 2$ $K$-matrices that give $SU(2)$ invariant theories can be obtained from the flux-trading procedure. Consider the matrices:

$$K = \begin{pmatrix} -m & -m-1 \\ -m-1 & -m \end{pmatrix} \tag{3.4}$$

These theories are also $SU(2)$ invariant since $K$ may be diagonalized as $K = \text{diag}(-2m-1, 1)$. While the $(1,1,2)$ Halperin state is not a spin singlet, a very similar state may be constructed (compare with (3.3)):

$$\Psi_{(-m,-m,-m-1)} = \det\left(\frac{\partial^{j-1}}{\partial w_i^{j-1}}\right) \det\left(\frac{\partial^{j-1}}{\partial z_i^{j-1}}\right) \Psi_{(m+1,m+1,m+1)}. \tag{3.5}$$

From a flux-trading standpoint, one has added $m+1$ flux tubes, but *antiparallel* to the magnetic field. The states (3.5) are the same as those introduced by Wu, Dev, and Jain [7]. These states fill in the odd-denominator spin-singlet states that are missing from the Halperin sequence, ie. the $(m, m, m-1)$ and $(-m, -m, -m-1)$ states cover all of the fractions $\nu = 2/p$ with $p$ odd.

It is clear that all states obtained from similar constructions by adding flux to $\nu = 2N$ spin-singlet states with $2N$ filled Landau levels are spin-singlets (in fact, $SU(2N)$ singlets). These states, at filling fractions $\nu = \frac{2N}{4Nk\pm 1}$, have $2N \times 2N$ $K$-matrices which have $2k+1$'s along the diagonal and $2k$'s off-diagonal, or $-2k+1$'s along the diagonal and $-2k$'s off-diagonal. These states may all be obtained from the hierarchy construction, and are the simplest spin-singlets in the hierarchy. There are others, however. These may be identified, using the machinery developed so far, by transforming to the symmetric basis. For example, let us find, at the second level of the hierarchy, those $(p, p, q)$ daughters of $(m, m, n)$ parent states which are overall spin-singlets. Such states will have hierarchical basis $K$-matrices of the form:

$$K^h = \begin{pmatrix} -m & -n & 1 & 0 \\ -n & -m & 0 & 1 \\ 1 & 0 & -p & -q \\ 0 & 1 & -q & -p \end{pmatrix} \qquad (3.6)$$

for a state in which quasiholes condense, and

$$K^h = \begin{pmatrix} m & n & 1 & 0 \\ n & m & 0 & 1 \\ 1 & 0 & -p & -q \\ 0 & 1 & -q & -p \end{pmatrix} \qquad (3.7)$$

for a state in which quasiparticles condense.

Recall that in the hierarchy basis the electromagnetic coupling is only to the first components. One can pass from this to the the symmetric basis, where the electromagnetic coupling is the same to all components. Then we find instead:

$$K^s = \begin{pmatrix} -m & -n & -m+1 & -n \\ -n & -m & -n & -m+1 \\ -m+1 & -n & 2-p-m & -q-n \\ -n & -m+1 & -q-n & 2-p-m \end{pmatrix} \qquad (3.8)$$

and

$$K^s = \begin{pmatrix} m & n & m+1 & n \\ n & m & n & m+1 \\ m+1 & n & 2+m-p & n-q \\ n & m+1 & n-q & 2+m-p \end{pmatrix} \qquad (3.9)$$

Now by demanding that $(1, -1, 1, -1)$ should be an eigenvector with unit eigenvalue we arrive at the conditions $m = n \pm 1$ in the respective cases, and the common condition $q = p - 2$. In other words, the parent state must be a spin singlet, $(m, m, m-1)$ or $(m-1, m-1, m)$, and the daughter state must be of the form $(p, p, p-2)$.

The Jain construction states at the second level of the hierarchy are $(2, 2, 0)$ daughter states of spin-singlet parents. The experimentally observed [2] $\nu = 4/3$ spin-singlet state which is the particle-hole conjugate of the $\nu = 2/3$ spin-singlet is a $(0, 0, -2)$ daughter state of quasiholes built on the $(1, 1, 0)$ state:

$$K^h = \begin{pmatrix} -1 & 0 & 1 & 0 \\ 0 & -1 & 0 & 1 \\ 1 & 0 & 0 & 2 \\ 0 & 1 & 2 & 0 \end{pmatrix} \tag{3.10}$$

Many states enjoy hidden symmetries not associated with spin. Indeed, as we mentioned earlier, even some standard hierarchy states of the one-layer spin-polarized hierarchy can have such symmetries. For instance, the spin-polarized $\nu = 2/3$ state has $K$-matrix:

$$K^h = \begin{pmatrix} -1 & -1 \\ -1 & 2 \end{pmatrix} \equiv K^s = \begin{pmatrix} -1 & -2 \\ -2 & -1 \end{pmatrix} \tag{3.11}$$

where $\equiv$ denotes $SL(\kappa, Z)$ equivalence. At the $N^{th}$ level of the hierarchy, there are states with $SU(N)$ symmetry – namely those at $\nu = \frac{N}{2kN \pm 1}$, which are among the most prominent experimentally (these are the states emphasized by Jain [8]). It is tempting to speculate that the higher symmetry may be energetically advantageous.

# 4. The Phase Transition at $\nu = 2/3$

It is quite interesting that the $K$-matrices are equivalent for the spin-singlet and spin-polarized states at $\nu = 2/3$. This fact has a simple physical interpretation. Both of these states are related by flux-trading to $\nu = 2$ states; in one case to the spin-singlet state, in the other, to the spin-polarized state. At $\nu = 2$, the lowest Landau level with spins aligned along the magnetic field is fully occupied. Then, depending on the ratio of the Zeeman and cyclotron energies, either the second spin-aligned Landau level or the first spin-reversed Landau level is filled (see Figure 1). A transition between these states is a simple level-crossing, and, hence, a first-order phase transition. Both states have $SU(2)$ symmetries. In one case, it is a symmetry between two Landau levels of the same spin. In the other case it is between two Landau levels of opposite spin.

At $\nu = 2$, the Zeeman energy is typically much smaller than the cyclotron energy, so the spin-singlet is realized. At $\nu = 2/3$, however, the gauge field fluctuations that must be accounted for in flux-trading arguments cause the effective mass to be renormalized and Zeeman and cyclotron energies are of roughly the same order of magnitude. Hence, the transition between these states may be seen at reasonable angles in tilted-field experiments which increase the Zeeman energy while holding the cyclotron energy constant. Since the spin-singlet and spin-polarized states have equivalent $K$-matrices at all of the fractions $\nu = \frac{2}{2m-1}$, a similar transition is kinematically allowed at all of these filling fractions.

At the first-order phase transition point, we expect phase coexistence. At the boundaries between spin-polarized and spin-singlet regions, there will be edge modes. These edges are a slightly more complicated version of the familiar edges which occur at the boundary between a Hall fluid and "vacuum". Here we have an edge between two Hall states. Such an edge will have modes traveling in both directions which could pair off and form a gap. At the notional edge between two Hall fluids which are the same, for instance, gap formation will occur unless the coupling between oppositely directed modes (the non-universal $V_{ij}$'s) is very small, which corresponds to weakly coupled fluids. This seems quite reasonable, because the boundary between two Hall fluids which are the same is not really a boundary at all.

Now at the boundary between our spin-polarized and spin-singlet regions the $K$-matrices for the fluids on both sides of the boundary are the same, so at this level we encounter the same situation as that which we just considered. Since the $V_{ij}$'s coupling charged modes reflect the basic Coulomb interaction, they are generic and we are led to predict that the charge transfer gap never closes. The $V_{ij}$'s which couple oppositely moving neutral modes are less robust, so further analysis is required to determine whether or not these are gapless.

## 5. Comments

1. These considerations suggest a construction for trial wavefunctions, as follows. Consider the $(m, m, m-1)$ spin-singlet states of electrons:

$$\Psi_{(m,m,m-1)}(z_i, \sigma_i) = \prod_{i<j}(z_i - z_j)^{K(\sigma_i, \sigma_j)} \tag{5.1}$$

These states can be represented by density matrices satisfying $\rho^2 = \rho$ (the condition for a pure state), where

$$\rho(z_i, \sigma_i; z_i', \sigma_i') = \Psi^*_{(m,m,m-1)}(z_i, \sigma_i)\Psi_{(m,m,m-1)}(z_i', \sigma_i') \tag{5.2}$$

Since we are working in the lowest Landau level, we can take a discrete basis and work with finite matrices, $\rho(I_i, \sigma_i; I_i', \sigma_i')$, where the $I_i$ label states in the lowest Landau level. If we now integrate out the spins,

$$\tilde{\rho}(I_i, I_i') = \sum_{\sigma_i} \rho(I_i, \sigma_i; I_i', \sigma_i) \tag{5.3}$$

we will produce a density matrix which, in general, describes a mixed state. If (as we expect) $\tilde{\rho}$ is dominated by a single pure state – that is, it has low entropy – then this state is a candidate to describe the *spin-polarized* state at $\nu = \frac{2}{2m-1}$. By extension, we can produce similar candidate states at $\nu = \frac{N}{2kN-1}$.

The key issue, of course, is whether $\tilde{\rho}$ is dominated by a single state. It is not implausible since the first factor on the right-hand-side of $\Psi_{(m,m,m-1)} = \Psi_{(m-1,m-1,m-1)} \times \Psi_{(1,1,0)}$ – which is dominant at large $m$ – is unaffected by the sum over spins. This argument is reminiscent of Jain's original argument [8], with the selection of a single pure state from a mixed ensemble playing the role of projection into the lowest Landau level. The philosophy here is that a given $K$-matrix encodes the electron-electron correlations of a universality class of quantum Hall states; if the wavefunctions for one particular realization in this class are easy to construct, we should try to use them to construct wavefunctions for other states in this class.

2. As we have noted, a great many Hall states have large symmetries hidden in their $K$-matrices. Unfortunately these symmetries are of limited use for the analysis of edge states, unless the symmetry is fundamental, because the edge theories are determined not only by the $K$-matrices, but also by a non-universal interaction matrix $V_{ij}$ [9]:

$$L = K_{ij}\partial_t\phi_i\partial_x\phi_j - V_{ij}\partial_x\phi_i\partial_x\phi_j \qquad (5.4)$$

If the symmetry of $K$ does not reflect a fundamental symmetry of the system, then it will generically be broken by $V_{ij}$. One might attempt to emphasize the importance of $K$ as against $V$ in layered samples by having a sharp field gradient at the edge, or by having the largest possible separations between the layers.

Symmetry-breaking terms in the bulk are less relevant (in a renormalization-group sense) than the Chern-Simons terms, so the symmetry is respected at low energies. These symmetries provide selection rules, just as in the case of atomic transitions. In particular, scattering of quasiholes and quasiparticles must conserve $SU(N)$ quantum numbers[*].

3. It is an important problem for string theory, to relate the field-theoretic paradigm of symmetry selection by space-time condensates to world-sheet phenomena. The symmetry selection mechanism described here has at least some of this flavor: the underlying Hall state is determined by a condensate of "vacuum" vertex operator (dressed electron) insertions, and its correlations determine the symmetry classification of non-vacuum (quasi-particle) insertions.

4. There are reasons to think that gauge field interactions depending in a non-trivial way upon spin can be useful in describing the anomalous correlated behavior of electrons in the CuO materials. It has been difficult, however, to see how such interactions could respect the observed spin-singlet character of many of the interesting phases. Above, we have seen that in theories of this type the symmetries can easily be present, but hidden.

---

[*] The creation of quasiparticle-quasihole pairs, however, is a *high-energy* process, which need not conserve $SU(N)$

Figure Caption

Figure 1. The single particle energy levels when the Zeeman energy is smaller than, but comparable to, the cyclotron energy. By turning on a magnetic field in the plane, the spin-reversed levels can be raised, resulting in a transition, at $\nu = 2$, from a spin-singlet to a spin-polarized state. The states at $\nu = 2/3$ are related to these states by adiabatic trading of statistical and magnetic flux.

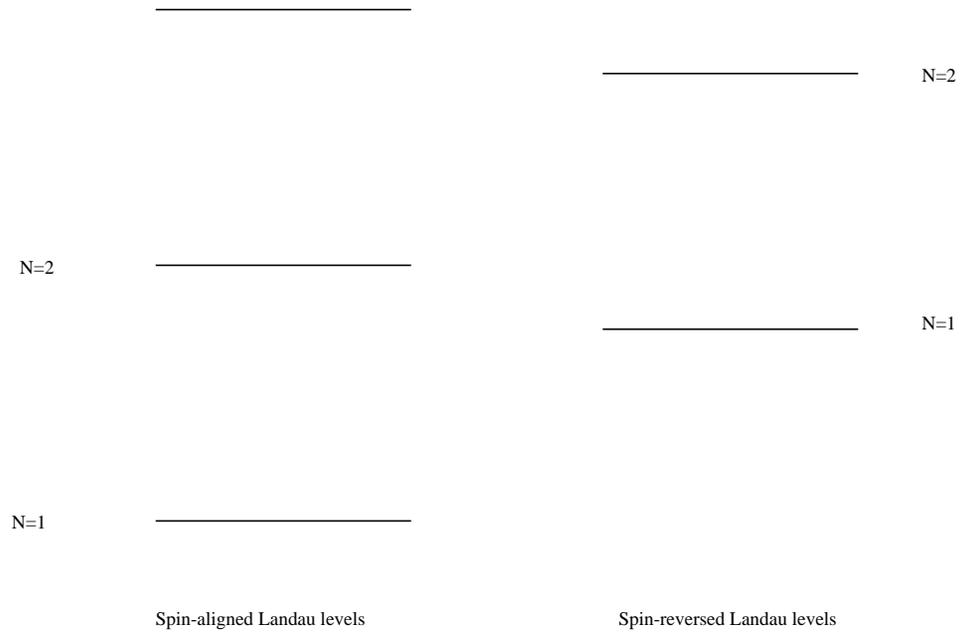

Figure 1.